\documentclass[twocolumn,showpacs,prl]{revtex4}

\usepackage{graphicx}
\usepackage{dcolumn}
\usepackage{bm}

\begin{document}

\title{Nonmonotonic $d_{x^2-y^2}$ superconducting gap in
electron-doped Pr$_{0.89}$LaCe$_{0.11}$CuO$_4$: Evidence of coexisting
antiferromagnetism and superconductivity?}

\author{Tanmoy Das, R. S. Markiewicz and A. Bansil}
\address{Physics Department, Northeastern University, Boston MA 02115,
USA}
\begin{abstract}

  Recent experiments on Pr$_{0.89}$LaCe$_{0.11}$CuO$_4$ observe an
  anisotropic spin-correlation gap and a nonmonotonic superconducting
  (SC) gap, which we analyze within the framework of a
  $t-t^{\prime}-t^{\prime\prime}-t^{\prime\prime\prime}-t^{iv}-U$
  model with a $d_{x^2-y^2}$ pairing interaction including a third
  harmonic contribution. By introducing a realistic broadening of the
  quasiparticle spectrum to reflect small-angle scattering, our
  computations explain the experimental observations, especially the
  presence of a maximum in the leading edge gap in the vicinity of the
  hot-spots. Our analysis suggests that the material behaves like a
  {\it two-band} superconductor with the $d$-wave third harmonic
  acting as the {\it interband pairing gap}, and that the
  anti-ferromagnetic (AFM) and SC orders co-exist in a uniform phase.

\end{abstract}
\maketitle
\narrowtext

Recent Raman scattering\cite{raman} and ARPES experiments\cite{plcco}
on electron-doped Pr$_{0.89}$LaCe$_{0.11}$CuO$_4$ (PLCCO) find a
$d_{x^2-y^2}$ superconducting (DSC) gap, which varies
non-monotonically along the Fermi surface (FS). The gap increases as
one moves away from the nodal direction towards the hot-spot region in
the Brillouin zone (BZ) where it attains a maximum value. It then
decreases as one approaches the zone boundary along the anti-nodal
direction. The involvement of the hot-spots suggests that the bosonic
pairing possesses a magnetic origin.\cite{gapmarkie,KYZK} If we assume
that the system continues to remain uniform with doping, and recall
that the residual gap in a half-filled AFM insulator yields two FS
pockets on electron doping, these observations present a striking
conundrum: The leading-edge gap is the largest in the momentum region
of the hot-spots where there are no segments of the FS and
quasiparticle states lie well below the Fermi energy.

In this article, we discuss a relatively simple route for resolving
this dilemma and understanding the behavior of the superconducting
state of PLCCO. In particular, we consider a uniformly doped system
with co-existing AFM and DSC orders, where the quasiparticle spectrum
is broadened realistically to reflect the effects of small angle
scattering.  We thus obtain in the normal state a finite spectral
weight at the Fermi energy ($E_F$) throughout the BZ, and when a third
harmonic term is included in the pairing interaction, the computed
leading edge gap along the FS reproduces the corresponding
experimental variations remarkably well. The third harmonic
contribution possesses the proper symmetry to couple the DSC order
parameter on the two FS pockets to induce a maximum in the leading
edge gap around the hot-spots even though the hot-spots lie outside
the momentum region of the two FS pockets. Notably, non-monotonic gap
variations have been obtained in some earlier
calculations\cite{gapmarkie,KYZK}, but we are not aware of a previous
study involving a realistic model of electron doping with two FS
pockets.

Our study gives insight into the issue of well known asymmetry between
the properties of cuprates with electron {\it vs} hole doping, which
has been a subject of considerable recent debate (see e.g.,
Ref.~\onlinecite{estripe}). There is growing evidence that the {\it
normal} state of the electron doped cuprates can be described as an
AFM metal up to a quantum critical point near optimal doping, and that
nano-scale phase separations or stripe physics so prominent in the
hole-doped case are weak or absent in the electron-doped
systems.\cite{kusko,Greene} Since our analysis based on a single phase
model in which the AFM and DSC orders co-exist is able to reasonably
explain non-monotonic gap variations, it supports the scenario that
the system remains uniform with electron doping not only in the normal
but also the SC state without the intervention of other orders.

We model PLCCO as a uniformly doped spin density wave (SDW)
antiferromagnet with $d$-wave superconductivity.  At the mean field
level, there is long-range AFM order, but when fluctuations are
included, the Neel temperature $T_N$ becomes zero (in the absence of
interlayer coupling), and the mean-field gap turns into a pseudogap
$\Delta^*$ with crossover temperature $T^*$. In the SC doping range
the FS consists of a necklace of two types of pockets: an
electron-like pocket near the antinodal point $(\pi ,0)$, and a
hole-like pocket near the $(\pi /2,\pi /2)$ nodal point. A pseudogap
near the hot-spot separates the two pockets. Our one-band Hubbard
model Hamiltonian is
\begin{eqnarray}\label{Ham}
H&=&\sum_{\vec{k},\sigma}\xi_{\vec{k}}c^{\dag}_{\vec{k},\sigma}c_{\vec{k},\sigma}+U_Q\sum_{\vec{k},\vec{k^{\prime}}}c^{\dag}_{\vec{k}+\vec{Q},\uparrow}c_{\vec{k},\uparrow}c^{\dag}_{\vec{k^{\prime}}+\vec{Q},\downarrow}c_{\vec{k^{\prime}},\downarrow}\nonumber\\
&&+\sum_{\vec{k},\vec{k^{\prime}}}V(\vec{k},\vec{k^{\prime}})c^{\dag}_{\vec{k},\uparrow}
c^{\dag}_{-\vec{k},\downarrow}c_{-\vec{k^{\prime}},\downarrow}c_{\vec{k^{\prime}},\uparrow}
\end{eqnarray}
where $c^{\dag}_{\vec{k},\sigma} (c_{\vec{k},\sigma})$ is the
electronic creation (destruction) operator with momentum $\vec{k}$ and
spin $\sigma$. The independent particle dispersion with respect to the
chemical potential $E_F$ is given by\cite{lin,foot2}
\begin{eqnarray}\label{dispersion}
\xi_{\vec{k}} &=& -2t[c_x(a) +c_y(a)] - 4t^{\prime}c_x(a)c_y(a) -
2t^{\prime\prime}[c_x(2a)\nonumber\\
&& + c_y(2a)] -4t^{\prime\prime\prime}[c_x(2a)c_y(a) +
c_y(2a)c_x(a)]\nonumber\\
&&-4t^{iv}c_x(2a)c_y(2a)-E_F,
\end{eqnarray}
with $c_i(\alpha a)$ =
cos$(\alpha \vec{k_i} a)$ and $a$ is the lattice constant. 

The effective on-site Hubbard repulsion $U_Q$ is taken to be
doping-dependent\cite{kusko}. The Mott gap $U_QS$ arises from a finite
expectation value of the staggered magnetization $S$ at the
commensurate ordering wave vector $\vec{Q} = (\pi,\pi)$. The $d$-wave
superconductivity including first and third harmonics is defined
by\cite{zaira,gapmarkie}
\begin{eqnarray}\label{delta}
\Delta_{\vec{k}}&=&\Delta_{1\vec{k}}
+\Delta_{3\vec{k}}=\sum_{\vec{k}^{\prime}}V(\vec{k},\vec{k^{\prime}})\langle
c_{\vec{k^{\prime}},\uparrow}^{\dag}c_{-\vec{k^{\prime}},\downarrow}^{\dag}\rangle \nonumber\\
&=&\sum_{i}\Delta_{i}g_{i\vec{k}} =\sum_i
g_{i\vec{k}}V_{i}\sum_{\vec{k^{\prime}}}g_{i\vec{k^{\prime}}}\langle
c_{\vec{k^{\prime}},\uparrow}^{\dag}c_{-\vec{k^{\prime}},\downarrow}^{\dag}\rangle,
\end{eqnarray}
with $i$ = 1,3 and $g_{i\vec{k}} = c_x(ia)-c_y(ia)$.

Hamiltonian of Eq.~1 is diagonalized straightforwardly
\cite{zaira}. The resulting quasiparticle dispersion consists of upper
and lower Hubbard bands (UHB and LHB), each gapped via the DSC pairing
as:
\begin{equation}\label{E}
E_{\vec{k}}^{U+(L+)} = -E_{\vec{k}}^{U-(L-)}=\sqrt{(\xi_{\vec{k}}^{+}\pm E_{0\vec{k}})^2 + \Delta_{\vec{k}}^2},
\end{equation}
where $E_{0\vec{k}}^2 = (\xi_{\vec{k}}^{-})^2 + (U_QS)^2$ and
$\xi_{\vec{k}}^{\pm} = (\xi_{\vec{k}} \pm \xi_{\vec{k}+\vec{Q}})/2$.
The SDW magnetization $S$ per site is evaluated self-consistently at
each temperature via the equations,
\begin{eqnarray}\label{S}
S&=& \sum_k \alpha_{\vec{k}}\beta_{\vec{k}}[[(v_{\vec{k}}^{-})^2 -(v_{\vec{k}}^{+})^2]+[(v_{\vec{k}}^{+})^2 -
(u_{\vec{k}}^{+})^2]f(E_{\vec{k}}^{U+})\nonumber\\
&&~~~~~~~~~ - [(v_{\vec{k}}^{-})^2 -(u_{\vec{k}}^{-})^2]f(E_{\vec{k}}^{L+})],
\end{eqnarray}
where $f(E) = 1/(\exp{(\beta E)}+1)$ is the fermi function with $\beta
= 1/k_BT$. The expressions for $\alpha_{\vec{k}}$, $\beta_{\vec{k}}$,
$u_{\vec{k}}^{\pm}$ and $v_{\vec{k}}^{\pm}$ are same as Eq (9) and Eq
(14) respectively of Ref.~\cite{zaira}. The self-consistent SC-gap
equations are
\begin{equation}\label{del}
\Delta_{i}=-V_{i}\sum_{\vec{k}}
\Delta_{\vec{k}}g_{i\vec{k}}\left[\frac{\tanh{(\beta
E_{\vec{k}}^{U+}/2)}}{2E_{\vec{k}}^{U+}} +\frac{\tanh{(\beta
E_{\vec{k}}^{L+}/2)}}{2E_{\vec{k}}^{L+}} \right]
\end{equation}

The one-particle Green function is
\begin{eqnarray}\label{green}
G(\vec{k},\omega) &=& \frac{[\omega Z +\xi_{\vec{k}}^{+}+E_{0\vec{k}}]f(E_{\vec{k}}^{U+})}{(\omega^2-\Delta_{\vec{k}}^2)Z^2 - (\xi_{\vec{k}}^{+} +E_{0\vec{k}})^2}\nonumber\\
&&+\frac{[\omega Z +\xi_{\vec{k}}^{+}-E_{0\vec{k}}]f(E_{\vec{k}}^{L+})}{(\omega^2-\Delta_{\vec{k}}^2)Z^2-(\xi_{\vec{k}}^{+} -E_{0\vec{k}})^2},
\end{eqnarray}
where we have included broadening due to small-angle elastic
scattering\cite{greenmarkie,greendahm} with the associated
renormalization factor, $Z=1+i\Gamma_0(\omega)\rm{sgn}(\omega)/
\sqrt{\omega^2 - \Delta_{\vec{k}}^2}$. $\Gamma_0(\omega)$ is the
normal state scattering rate, discussed below.  The corresponding
spectral intensity is $A(\vec{k},\omega)
=-\rm{Im}[G(\vec{k},\omega)]/\pi$.

We fitted the dispersion given by Eq.~\ref{E} to the ARPES results on
PLCCO\cite{plcco} at 30 K to obtain the relevant TB parameters as
follows: $t$ = 0.12 eV, $t^{\prime}$ = -0.06 eV, $t^{\prime\prime}$ =
0.034 eV, $t^{\prime\prime\prime}$ = 0.007 eV, $t^{iv}$ = 0.02 eV,
with $E_F$ = -0.082 eV at 11\% doping.\cite{foot1} The effective
Hubbard parameter is found to be $U_Q$ = 4.15$t$, which yields a
self-consistent value of the magnetization $S$ of 0.281 from
Eq.~\ref{S}.
\begin{figure}[tp]
\scalebox {0.4}{\includegraphics{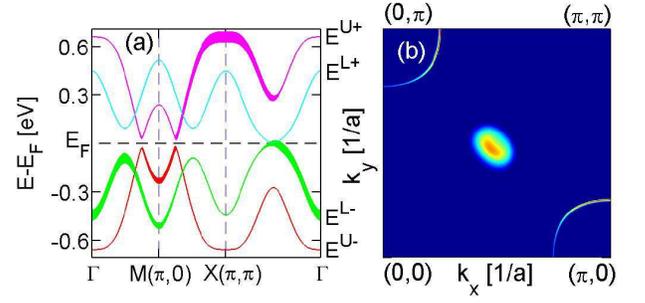}}
\caption{
  (color online) (a) Computed quasiparticle spectrum of PLCCO at 11\%
  doping. Artificially large values of the pairing interaction
  parameters ($\Delta_{1}$ = 20 meV and $\Delta_{3}$ = -6 meV) have
  been used to highlight the effect of the SC gap, which splits the
  UHB and the LHB into the pair of bands $E^{U\pm}$ (magenta and red)
  and $E^{L\pm}$ (cyan and green), respectively.  Spectral weight is
  represented by the vertical width of the bands.  (b) FS map in PLCCO
  at 30 K (normal state) obtained by integrating the spectral
  intensity of Eq. \ref{green} over a small energy window of $\pm$ 5
  meV around $E_F$. High intensity is denoted by red and low by
  blue. }
\label{bandfermi}
\end{figure}

Fig. 1 clarifies the nature of the quasiparticle spectrum and the FS.
As Eq. 4 shows, the non-interacting band is split via the Hubbard
interaction into the upper and lower Hubbard bands. As seen from
Fig. 1(a), each of these bands is further split by the SC interaction
to yield pairs of bands $E^{U\pm}$ (magenta and red) and $E^{L\pm}$
(cyan and green) which lie symmetrically above and below the
$E_F$. [Note that the SC gap has been made artificially large in
Fig. 1(a) to highlight the effect of the SC splitting.] The spectral
weights for various bands (proportional to the vertical width of
shading) are seen to vary greatly with {\bf k}. In effect, the
spectral weight of the non-interacting band is redistributed between
the UHB and the LHB through the AFM order, and it is modified further
via the SC order at very small energy scales of a few meV's.  Above
the $E_F$, we see from Fig. 1(a) that most of the spectral intensity
resides in the UHB branch $E^{U+}$ (magenta) along the
$(\pi,0)\longrightarrow (\pi,\pi) \longrightarrow (\pi/2,\pi/2)$ line.
At other momenta, the spectral weight generally lies below $E_F$ and
is carried by the $E^{L-}$ band (green). Fig.~\ref{bandfermi}(b) shows
a map of the normal state (30K) FS. This FS is consistent with the
experimental results on electron-doped Nd$_{0.87}$Ce$_{0.13}$CuO$_4$
(NCCO)\cite{nparm,ncco}, and it can be understood with reference to
the band structure of Fig. 1(a). The $E^{U-}$ band (red) gives rise to
the ($\pi$,0)-centered electron pockets, while the
$(\pi/2,\pi/2)$-centered hole pocket arises from the $E^{L-}$ band
(green).

\begin{figure}[tp]
\scalebox {0.46}{\includegraphics{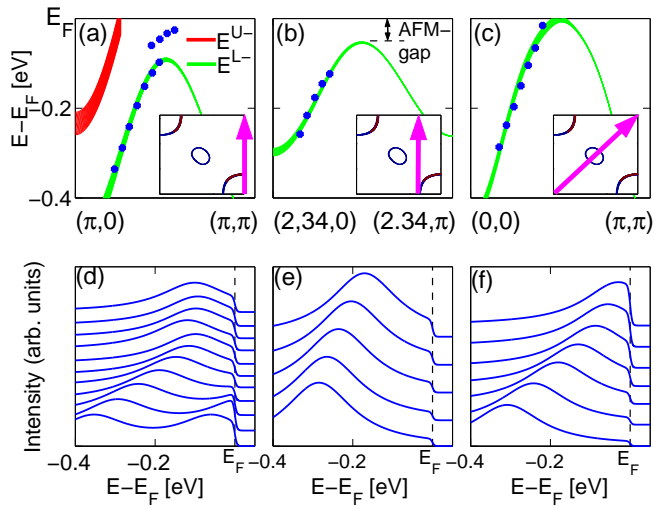}}
\caption{
  (color online) (a)-(c) Quasiparticle spectrum of PLCCO with 11\%
  doping in the {\it normal} state at 30 K along three different lines
  (see insets) in the $(k_x,k_y)$-plane. Spectral weights are
  proportional to the widths of lines as in Fig. 1(a). Blue dots give
  the corresponding experimental dispersions taken from
  Ref.\protect\onlinecite{plcco}. (d)-(f) Spectral intensity obtained
  from Eq.~\protect\ref{green} as a function of binding energy for a
  series of momenta where the actual momentum values correspond to the
  experimental points (blue dots) in (a)-(c).  } \label{fitinty}
\end{figure}

Fig.~\ref{fitinty} gives further insight into the nature of the
quasiparticle spectrum and the associated spectral intensity in PLCCO.
The computed energy bands in the normal state along the three cuts
used in the fitting procedure are compared directly with the
corresponding experimental dispersions in panels (a)-(c). The overall
agreement is seen to be quite good, although a discrepancy is evident
in (a) for the anti-nodal cut in that the computed size of the
electron pocket around ($\pi$,0) is smaller than that indicated by the
experimental data.\cite{foot3} Notably, due to the presence of the AFM
gap, quasiparticle states along the hot spot in (b) lie well below the
$E_F$, leading to a suppression of the spectral weight up to binding
energies of about 60 meV.  In contrast, along the nodal direction in
(c), the quasiparticle band crosses $E_F$, and as already pointed out,
gives rise to the $(\pi/2,\pi/2)$ centered hole pockets.

The spectral intensity computed from the imaginary part of the
one-particle Green's function of Eq.~7 is shown for a series of
momenta in the bottom row of Fig. 2. Here the cuts in the top and
bottom set of panels correspond to each other. For example, in (e) the
five spectra shown refer to the five k-points given by the blue dots
in (b) with the lowest spectrum in (e) corresponding to the leftmost
dot in (b). The computations assume a $k$-independent broadening
function of form: $\Gamma_0(\omega) = C_0[1 + (\omega/\omega_0)^{p}]$,
which is similar to a form which has been proposed for hole-doped
cuprates\cite{rice}. Parameter values of $C_0$ = 0.1 eV, $\omega_0$ =
1.59 eV, with $p$ =3/2 reasonably reproduce the experimentally
observed broadenings\cite{foot4}.  We emphasize that, for our
purposes, the detailed spectral shape is not so important.  The key is
the presence of quasiparticle broadening, which allows the development
of a finite spectral weight at the $E_F$ and the formation of the
leading edge superconducting gap at all momenta, even though the
underlying quasiparticle states lie well below the $E_F$ at most
momenta.

\begin{figure}[tp]
\scalebox{0.5}{\includegraphics{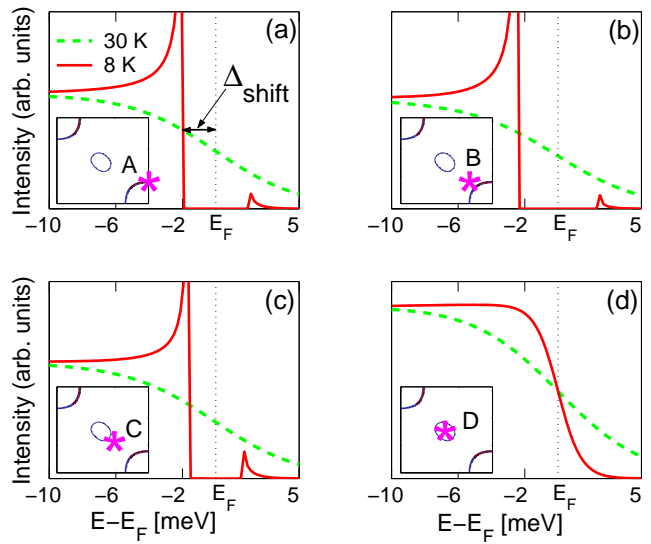}}
\caption{
  (color online) Spectral intensity computed from Eq. \ref{green} in
  the SC state at 8 K (solid lines) is compared with that for the
  normal state at 30 K (dashed lines) at four representative momentum
  points A-D (indicated by stars in the insets). Vertical dotted lines
  mark the $E_F$, which corresponds to the energy where the normal
  state spectral weight falls to half of its value at higher
  energy. The shift ($\Delta_{\rm{shift}}$) of the spectral intensity
  away from the $E_F$ as a leading edge gap opens up in the SC state
  is seen in (a)-(c), while this shift vanishes in (d) along the nodal
  direction due to the d-wave symmetry of the gap. }
\label{nef_inty_angle}
\end{figure}

Fig. 3 compares spectral intensities in the vicinity of the $E_F$ in
the normal and the SC state at four different momenta A-D (see
insets). We see that in (a)-(c) the midpoint of the leading-edge of
the 8 K spectrum (solid lines) is shifted by $\Delta_{\rm{shift}}$
towards higher binding energies with respect to the spectrum at 30 K
(dashed lines), while along the nodal direction in (d), this shift
vanishes due to the node in the d-wave gap.  Experiments\cite{plcco}
find a leading edge gap $\Delta_{\rm{shift}}$ of 2.0 meV along the
antinodal direction at the momentum point A ($\pi$,0.8), which
increases to 2.5 meV at the hotspot B(2.34,1.0); $\Delta_{\rm{shift}}$
then decreases to 1.6 meV at C(1.85,1.23) and becomes zero along the
nodal direction at the point D(1.52,1.52). The computations of Fig. 3
reproduce this behavior. For this purpose, the superconducting
parameters are found to be: $V_{1}$ = -76 meV and $V_{3}$ = 54 meV,
leading to self-consistent values of the gap parameters at 8 K of
$\Delta_{1}$ = 1.25 meV and $\Delta_{3}$ = -0.53 meV from
Eq.~\ref{delta} and \ref{del}. These parameter values yield $T_c$= 15
K, which is in reasonable accord with the experimental value of 26
K. Note that a leading edge gap is clearly seen near the hotspot in
(b). This gap is a direct consequence of the broadening of the
spectrum due to small angle scattering and the associated residual
spectral weight at the $E_F$, even though this spectral weight is
quite small due to the presence of a sizable pseudogap in the
spectrum.

Fig.~4(a) shows variations in the leading edge shift
$\Delta_{\rm{shift}}$ and its non-monotonic and anisotropic nature
more clearly. Here the {\it minimum} computed value of
$\Delta_{\rm{shift}}$ (blue line) is plotted along various directions
given by the angle $\phi$, where $\phi=0$ refers to the antinodal
direction and $\phi=45^o$ to the nodal direction. The actual momentum
points at which the computed values of $\Delta_{\rm{shift}}$ are
plotted, shown in the inset, lie close to the non-interacting FS. The
corresponding experimental values (red dots) are in good accord with
the theoretical results. In particular, the gap reaches a maximum
value of about 2.5 meV around $\phi=18^o$ near the hot-spot in both
theory and experiment. The monotonic d-wave gap obtained by setting
$\Delta_3 = 0$ in the calculations is also shown for reference (dashed
line) to highlight the non-monotonic gap variations in the present
case.

Our explanation of nonmonotonic gap variations should be distinguished
sharply from the notion that one simply observes two different gaps in
the experiments, i.e. an SC gap along the pockets and an AFM gap near
the hotspots. This alternative picture suffers from the problem that
the AFM gap in the normal state of the cuprates is far too large, and
moreover, it is not clear how the leading edge gap comes about since
there are no states at the $E_F$ away from the region of the two small
FS pockets.  Interestingly, the two band model study of
Ref.~\onlinecite{twoband2} would require an anomalously small value of
$t$ to fit the experimental gap values in PLCCO.
\begin{figure}[tp]
\scalebox {0.4}{\includegraphics{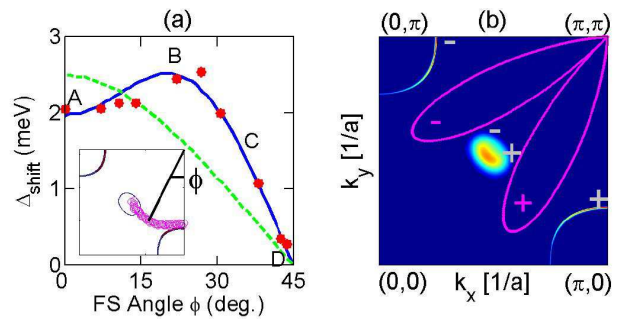}}
\caption{
  (color online) (a) Computed leading-edge gap (blue line)
  $\Delta_{\rm{shift}}$ at 8 K as a function of the FS angle $\phi$ is
  compared with the corresponding experimental results (red dots) on
  PLCCO\cite{plcco}. A-D are momentum points discussed previously in
  connection with Fig. 3. Green line gives the monotonic gap variation
  at 8 K in the absence of the third harmonic term (i.e.
  $\Delta_3=0$). The inset shows (open circles) the actual momentum
  points for various directions $\phi$ where the computed gap values
  are plotted. (b) Form of the third harmonic contribution
  $\Delta_{3\vec{k}}$ (magneta curve) to the gap (see Eq.~3) is shown
  schematically on top of the FS map of Fig. 1(b). } \label{del3}
\end{figure}

Third harmonic gaps have been found not only in electron-doped
PLCCO\cite{plcco} and NCCO\cite{raman}, but also in hole-doped
(underdoped) Bi$_2$Sr$_2$Ca$_{n-1}$Cu$_n$O$_{2n+4}$ ($n=1-3$)
(BISCO)\cite{bisco1,bisco2}. Interestingly, the third harmonic
contribution in electron-doped cuprates possesses an opposite sign and
its size is substantially larger than in the hole-doped case, leading
to the nonmonotonic gap variations discussed above. Our study provides
insight into this behavior. Separate electron and hole pockets in
PLCCO and NCCO are suggestive of two-band\cite{twoband1}
superconductivity as in the case of MgB$_2$ where interband pairing
greatly enhances the $T_c$\cite{SuMer}. However, the form of our
Hamiltonian in Eq. \ref{Ham} does not involve a separate interband
pairing gap.  Fig. \ref{del3}(b) shows that $\Delta_{3\vec{k}}$
possesses the correct symmetry and that its size is maximal in the
interval between the two pockets just like the leading edge gap
$\Delta_{\rm{shift}}$.  With this in mind, we propose that the
third-harmonic gap $\Delta_{3\vec{k}}$ can be looked upon as playing
the role of interband pairing gap in PLCCO and NCCO. In fact, the `hot
spot' between the two pockets is associated with strong AFM
fluctuations in a quasi-two-dimensional system\cite{greenmarkie,KYZK},
which in turn are responsible for coupling the UHB and the LHB, and
thus coupling the phase of the order parameter on the two FS segments.

The fact that the nonmonotonic gap in our model is intimately
associated with the AFM pseudogap indicates that superconductivity
arises in an AFM background. These results support the picture of
electron doped cuprates being uniformly doped AFM metals, even in the
superconducting state, rather than being phase separated AFM and SC
domains. Note however that when fluctuations are added, the present
AFM gap becomes a short-range order pseudogap, with a possible Neel
ordering transition at lower temperatures due to interlayer coupling.
Open questions remain as to whether superconductivity coexists with
this residual 3D Neel order, or whether the onset of superconductivity
destroys either the 3D order, or even the 2D ($T$=0K) long-range AFM
order.

\begin{acknowledgments}

  We thank Hong Ding for sharing some of his unpublished results with
  us.  This work is supported by the U.S.D.O.E contract
  DE-AC03-76SF00098 and benefited from the allocation of supercomputer
  time at NERSC and Northeastern University's Advanced Scientific
  Computation Center (ASCC).

\end{acknowledgments}


\begin{thebibliography}{99}
\bibitem{foot2}
%
In analogy with NCCO, we expect interlayer coupling effects to be
small in PLCCO--see Ref~\onlinecite{lin}.
\bibitem{foot1}
%
Specifically, we fitted the experimental dispersions along three
different directions given in Figs 2 (d)-(f) of
Ref.~\onlinecite{plcco}.
\bibitem{foot3}
%
If we attempt to fit a larger electron pocket, we obtain an
inconsistently larger value of the doping level than the 11\% value
appropriate for the PLCCO sample used in the experiments of
Ref.~\onlinecite{plcco}.
\end{thebibliography}
\end{document}